\documentclass[11pt,a4paper]{article}
 
\usepackage{amsmath,amsthm,amssymb}

\usepackage{cite}

\pdfoutput=1

\usepackage{graphics,graphicx}
\usepackage{epsfig}
\usepackage{multicol}
\usepackage{color}
\makeatletter
\@addtoreset{equation}{section}
\makeatother

\usepackage{braket}
\usepackage{bm}

\begin{document} 
\sloppy

\begin{center}
\LARGE{{\bf Problems with Modified Commutators}}
\end{center}

\begin{center}
\large{Matthew J. Lake${}^{a,b,c,d,e*}$\footnote{matthewjlake@narit.or.th} and Anucha Watcharapasorn${}^{b,f*}$\footnote{anucha@stanfordalumni.org}}
\end{center}
\begin{center}
\emph{$^{a}$National Astronomical Research Institute of Thailand, \\ 260 Moo 4, T. Donkaew,  A. Maerim, Chiang Mai 50180, Thailand \\}
\emph{$^{b}$Department of Physics and Materials Science, \\ Faculty of Science, Chiang Mai University, \\ 239 Huaykaew Road, T. Suthep, A. Muang, Chiang Mai 50200, Thailand \\}
\emph{$^{c}$School of Physics, Sun Yat-Sen University, \\ Guangzhou 510275, People’s Republic of China \\}
\emph{$^{d}$Department of Physics, Babe\c s-Bolyai University, \\ Mihail Kog\u alniceanu Street 1, 400084 Cluj-Napoca, Romania \\}
\emph{$^{e}$Office of Research Administration, Chiang Mai University, \\ 239 Huaykaew Rd, T. Suthep, A. Muang, Chiang Mai 50200, Thailand \\}
\emph{$^{f}$Center of Excellence in Quantum Technology, \\ Faculty of Engineering, Chiang Mai University, \\ 239 Huaykaew Rd, T. Suthep, A. Muang, Chiang Mai 50200, Thailand}
\vspace{0.1cm}
\end{center}

\begin{abstract}
The purpose of this paper is to challenge the existing paradigm on which contemporary models of generalised uncertainty relations (GURs) are based, that is, the assumption of modified commutation relations. 
We review an array of theoretical problems that arise in modified commutator models, including those that have been discussed in depth and others that have received comparatively little attention, or have not been considered at all in the existing literature, with the aim of stimulating discussion on these topics. 
We then show how an apparently simple assumption can solve, or, more precisely, evade these issues, by generating GURs without modifying the basic form of the canonical Heisenberg algebra. 
This simplicity is deceptive, however, as the necessary assumption is found to have huge implications for the quantisation of space-time and, therefore, gravity. 
These include the view that quantum space-time should be considered as a quantum reference frame (QRF) and, crucially, that the action scale characterising the quantum effects of gravity, $\beta$, must be many orders of magnitude smaller than Planck's constant, $\beta \sim 10^{-61} \times \hbar$, in order to recover the present day dark energy density. 
We argue that these proposals should be taken seriously, as a potential solution to the pathologies that plague minimum length models based on modified commutators, and that their implications should be explored as thoroughly as those of the existing paradigm, which has dominated research in this area for almost three decades. 
\end{abstract}

\section*{}
{{\bf Keywords}: Generalised uncertainty relations, generalised uncertainty principle, extended uncertainty principle, modified commutation relations, minimum length, minimum momentum, quantum geometry, quantum gravity}

\newpage

\tableofcontents 

\newpage

\section{Introduction} \label{Sec.1}

Thought experiments in quantum gravity suggest the existence of generalised uncertainty relations (GURs) \cite{Maggiore:1993rv,Adler:1999bu,Scardigli:1999jh,Bolen:2004sq,Park:2007az,Bambi:2007ty}
and two of the most widely studied GURs are known as the generalised uncertainty principle (GUP) and the extended uncertainty principle (EUP). 
These may be written as
\begin{eqnarray} \label{GUP-1}
\Delta x^i \gtrsim \frac{\hbar}{2\Delta p_j} \delta^{i}{}_{j} \left[1 + \alpha_0 \frac{2G}{\hbar c^3}(\Delta p_j)^2\right] \, ,
\end{eqnarray} 
and
\begin{eqnarray} \label{EUP-1}
\Delta p_j \gtrsim \frac{\hbar}{2\Delta x^i} \delta^{i}{}_{j} \left[1 + 2\eta_0 \Lambda (\Delta x^i)^2\right] \, , 
\end{eqnarray} 
respectively, where $\alpha_0$ and $\eta_0$ are numerical constants of order unity. 
The GUP implies the existence of a minimum length scale of the order of the Planck length \cite{Maggiore:1993rv,Adler:1999bu,Scardigli:1999jh}
whereas the EUP implies a minimum momentum scale of the order of the de Sitter momentum \cite{Bolen:2004sq,Park:2007az,Bambi:2007ty}
For later convenience, we define the Planck and de Sitter scales as 
\begin{eqnarray} \label{Planck-dS}
l_{\rm Pl} := \sqrt{\hbar G/c^3} \simeq 10^{-33} \, {\rm cm} \, , \quad m_{\rm Pl} := \sqrt{\hbar c/G} \simeq 10^{-5} \, {\rm g} \, , 
\nonumber\\
l_{\rm dS} := \sqrt{3/\Lambda} \simeq 10^{28} \, {\rm cm} \, , \quad m_{\rm dS} := (\hbar/c)\sqrt{\Lambda/3} \simeq 10^{-66} \, {\rm g} \, , 
\end{eqnarray} 
where $\Lambda \simeq 10^{-56} \, {\rm cm^{-2}}$ is the cosmological constant \cite{Betoule:2014frx,Aghanim:2018eyx}. 
Assuming both minimum length and momentum scales suggests the extended generalised uncertainty principle (EGUP), 
\begin{eqnarray} \label{EGUP-1}
\Delta x^i\Delta p_j \gtrsim \frac{\hbar}{2} \delta^{i}{}_{j} \left[1 + \alpha_0 \frac{2G}{\hbar c^3}(\Delta p_j)^2 + 2\eta_0\Lambda (\Delta x^i)^2\right] \, ,
\end{eqnarray} 
but the relations (\ref{GUP-1})-(\ref{EGUP-1}) are heuristic and it remains an open problem how to derive the GUP, EUP and EGUP rigorously, from a modified quantum formalism. 

Until recently, the only method considered in the existing literature was to modifiy the canonical commutation relations such that \cite{Tawfik:2014zca,Tawfik:2015rva}
\begin{eqnarray} \label{EGUP_[X,P]-1}
[\hat{x}^i,\hat{p}_j] = i\hbar\delta^{i}{}_{j} \ \hat{\mathbb{I}} \mapsto [\hat{X}^i,\hat{P}_j] = i\hbar\delta^{i}{}_{j} F(\hat{\bf{P}},\hat{\bf{X}}) \, ,
\end{eqnarray}
which gives rise to GURs via the Schr{\" o}dinger-Robertson relation \cite{Robertson:1929zz,Schrodinger:1930ty}
Throughout this work we use capital letters to denote modified operators, that give rise to modified commutators, and lower case letters to denote their canonical quantum counterparts. 
Unfortunately, this apparently reasonable assumption has been shown to give rise to a variety of pathologies \cite{Tawfik:2014zca,Hossenfelder:2012jw,Hossenfelder:2014ifa}. 
These strongly suggest that modified commutator models are not mathematically self-consistent \cite{Lake:2020rwc,Lake-Frontiers-1}. 

In this paper, we review six fundamental problems encountered by GUR models based on modified commutation relations:
\begin{enumerate}

\item violation of the equivalence principle,

\item violation of Lorentz invariance in the relativistic limit,

\item the `soccer ball' problem for multi-particle states,

\item the reference frame-dependence of the `minimum' length,

\item the background geometry is {\it not} quantum,

\item the mathematical inconsistency of modified phase space volumes. 

\end{enumerate}

The first three of these have been discussed at length in the literature (see, for example \cite{Tawfik:2014zca,Hossenfelder:2012jw,Hossenfelder:2014ifa} and references therein) and we review them only briefly. 
The fourth and fifth problems were discussed previously in \cite{Lake:2020rwc} but, to the best of our knowledge, have not been discussed elsewhere. 
The sixth and final problem raised in this short review has, surprisingly, not been considered before. 
Nonetheless, we argue that it represents the most serious objection yet raised against the modified commutator paradigm. 

We review each problem, sequentially, in Secs. \ref{Sec.2.1}-\ref{Sec.2.6}. 
In Sec. \ref{Sec.3}, we consider the relative importance of each, and ask whether or not such problems could instead be viewed as features, rather than bugs, of a viable extension of canonical quantum mechanics. 
An alternative model, that circumvents these issues without the use of modified commutation relations, is reviewed in Sec. \ref{Sec.4}. 
Our conclusions are summarised in Sec. \ref{Sec.5}.

\section{Problems with modified commutators} \label{Sec.2}

\subsection{Violation of the equivalence principle} \label{Sec.2.1}

In canonical quantum mechanics (QM), the Heisenberg equation for the time evolution of an arbitrary Hermitian operator $\hat{O}$ is 
\begin{eqnarray} \label{Heisenberg_Eq}
\frac{d}{dt}\hat{O}(t) = \frac{i}{\hbar}[\hat{H},\hat{O}] + \left(\frac{\partial\hat{O}}{\partial t}\right)_{H} \, ,
\end{eqnarray}
where $\hat{H} = \hat{p}^2/(2m) + V(\hat{\bf{x}})$. 
For the position operator $\hat{x}^{i}(t)$ this gives 
\begin{eqnarray} \label{x(t)}
\frac{d}{dt}\hat{x}^{i}(t) = \frac{\hat{p}^{i}}{m} \, ,
\end{eqnarray}
where right-hand side follows from the form of the canonical position-momentum commutator, $[\hat{x}^{i},\hat{p}_{j}] = i\hbar \delta^{i}{}_{j} \, \hat{\mathbb{I}}$. 
From (\ref{x(t)}), it follows that the acceleration of the position expectation value of a quantum particle is independent of its mass:
\begin{eqnarray} \label{}
\hat{a}^{i} = \frac{1}{m}\frac{d\hat{p}^{i}}{dt} = \frac{d^2\hat{x}^{i}}{dt^2} \, .
\end{eqnarray}

For the generalised operators $\hat{X}^{i}$ and $\hat{P}^{j}$ satisfying the modified commutator
\begin{eqnarray} \label{[X,P]-1}
[\hat{X}^i,\hat{P}_j] = i\hbar\delta^{i}{}_{j}G(\hat{\bf{P}}) \, , 
\end{eqnarray}
the Heisenberg equation for $\hat{X}^{i}(t)$ is 
\begin{eqnarray} \label{X(t)}
\frac{d}{dt}\hat{X}^{i}(t) = \frac{\hat{P}^{i}}{m}G(\hat{\bf{P}}) \, ,
\end{eqnarray}
so that, for $G(\hat{\bf{P}}) \neq 1$, the particle experiences a mass-dependent acceleration:
\begin{eqnarray} \label{F^i}
\hat{A}^{i} = \frac{1}{m}\frac{d\hat{P}^{i}}{dt} = \frac{1}{G(\hat{\bf{P}})}\left[\frac{d^2\hat{X}^{i}}{dt^2} - \frac{\hat{P}^{i}}{m}\frac{dG(\hat{\bf{P}})}{dt}\right] \, .
\end{eqnarray}
Although there is no universally agreed upon formulation of the equivalence principle (EP) for quantum systems \cite{Paunkovic:2022flx}, it is clear that such acceleration violates any sensible definition of the EP in the quantum regime, and, crucially, no experimental evidence has yet been found to support its existence. 

This argument assumes that the generalised Hamiltonian takes the form $\hat{H} = \hat{P}^2/(2m) + V(\hat{\bf{X}})$ and that a well defined Heisenberg picture exists in the generalised theory, but both of these are reasonable assumptions. 
Similar analyses demonstrate that the EP is also violated in models with modified commutators characterised by the functions $G(\hat{\bf{X}})$ and $G(\hat{\bf{X}},\hat{\bf{P}})$. 
It is therefore impossible to obtain the GUP, EUP, or EGUP from modified commutator models without violating the founding principles of classical gravity and, most likely, any viable generalisation of the these principles that includes the quantum realm \cite{Paunkovic:2022flx}. 

\subsection{Violation of Lorentz invariance} \label{Sec.2.2}

We recall that the canonical Heisenberg algebra is simply an $\hbar$-scaled representation of the shift-isometry algebra of Euclidean space and of space-like slices of flat space-time in the relativistic limit, i.e., it is the the translation subgroup of the full Poincar{\' e} group that characterises the symmetries of Minkowski space, including translations, Lorentz boosts and rotations. 
Therefore, any modification of the Heisenberg algebra implies the violation of translational symmetry {\it unless} we choose to interpret it as a manifestation of a modified de Broglie relation. 
In this case, the physical momentum $\bf{p}$ is a nonlinear function of the wavenumber $\bf{k}$, but the latter may still be identified with the shift-isometry generator of the background space, $\hat{\bf{k}} \equiv \hat{\rm{d}}_{\bf{x}}$. 

Unfortunately, both these scenarios lead to inconsistencies. 
In the first, in which we interpret the modified Heisenberg algebra as a manifestation of broken translation invariance, one faces a problem in defining the classical limit of the theory. 
Implementing a canonical quantisation scheme $\left\{O_1,O_2\right\}_{\rm PB} = \rm{lim}_{\hbar \rightarrow 0}\frac{1}{i\hbar}[\hat{O}_1,\hat{O}_2]$ and requiring the correspondence principle \cite{Rae} to hold implies an equivalent modification of the canonical Poisson brackets. 
This violates Galilean invariance, even for classical macroscopic systems, and, hence, Poincar{\' e} invariance in the relativistic limit. 
To date, no evidence for the breaking of Poincar{\' e} invariance, including shift invariance, has been found, although bounds on the symmetry breaking parameters have been determined from a wide range of experiments \cite{Gupta:2022qoq}. 

In the second scenario, one encounters problems related to the nonlinearity of $p(k)$, where $p = (p_0,-\bf{p})$ and $k =  (k_0,-\bf{k})$ denote the relativistic 4-momentum and its corresponding wave number, respectively. 
(From here on, we neglect space-time indices for the sake of notational convenience.) 
When $p(k)$ is nonlinear it is unclear if we should require the physical momentum $p$ or wave number $k$, also known as the pseudo-momentum, to transform under the Poincar{\' e} group. 
Choosing the wave number as the Lorentz invariant quantity, the Lorentz transformations become nonlinear functions of $k$ and the transformation of the sum $k_1 + k_2$ is no longer equal to the sum of the transformations of $k_1$ and $k_2$, individually. 
Likewise, choosing $p$ as the Lorentz invariant variable, which is physically more reasonable, a similar problem occurs and the transformation of $p_1 + p_2$ is no longer equal to the sum of the individual transformations of $p_1$ and $p_2$. 
Each case requires the definition of a new nonlinear addition law, either for the pseudo-momenta, or for the physical momenta, respectively \cite{Hossenfelder:2012jw}.

In the latter case, the new sum rule for the physical momenta is independent of the chosen inertial frame, by construction, but a new problem is created. 
If the nonlinear composition function has a maximum at the Planck momentum, as implied by consistency with the GURs generated by the modified de Broglie relation, then the sum of momenta will never exceed this maximum value. 
The Planck momentum, $m_{\rm Pl}c \simeq 10^{5} \, {\rm g} \, . \, {\rm cm s}^{-1}$, is large for fundamental particles with rest masses $m \ll m_{\rm Pl}$ but very small for macroscopic objects with rest masses $M \gg m_{\rm Pl}$, which may easily exceed it at ordinary non-relativistic velocities. 
The problem of reproducing a sensible multi-particle limit when choosing the physical momentum to transform under modified (nonlinear) Lorentz transformations is known as the ‘soccer ball problem’ \cite{Hossenfelder:2012jw,Hossenfelder:2014ifa}. 
This will be considered in more detail in the following section, in which we outline a recently proposed solution \cite{Amelino-Camelia:2014gga}, and its critique.

From the remarks above it is clear that the introduction of nonlinear de Broglie relations $\bf{p(k)}$ in non-relativistic QM requires that $p = (p_0,-\bf{p})$ must be a nonlinear function of $k = (k_0,-\bf{k})$ in the relativistic limit. 
This makes Lorentz violation unavoidable {\it unless} one introduces a new nonlinear composition law, either for the pseudo-momentum $k$, or the physical momentum $p(k)$. 
However, this leads to new problems, and it is not clear whether sensible multi-particle limits of such theories exist \cite{Hossenfelder:2012jw,Hossenfelder:2014ifa}. 
In Sec. \ref{Sec.2.3} we argue that, despite valiant attempts \cite{Amelino-Camelia:2014gga}, the soccer ball problem has not, in fact, been solved. 
This shows that a sensible relativistic limit of an arbitrary GUR model cannot be obtained by introducing a nonlinear composition law for Lorentz boosts, and one is left with Lorentz violation as the only possible outcome of such theories. 
Though not absolutely ruled out experimentally, the parameters characterising such violations are severely constrained by observations \cite{PerezdelosHeros:2022izj}. 

Nonetheless, this does not necessarily mean that GURs imply Lorentz violation. 
The problem, here, is the derivation of GUP- and EUP-type relations from the assumption of a modified Heisenberg algebra. 
In Sec. \ref{Sec.3} we show how the GUP, EUP and EGUP can be derived from an alternative mathematical structure, which leaves the canonical Heisenberg algebra unchanged except for a simple rescaling of the form $\hbar \ \rightarrow \hbar + \beta$, and which, therefore, is compatible with Lorentz symmetry in the relativistic regime. 

\subsection{The soccer ball problem} \label{Sec.2.3}

A brief overview of the soccer ball problem was given in the previous section, in connection with the issue of Lorentz violation, and we will not repeat it here. 
Instead, we focus on the main proposal for a solution of the problem \cite{Amelino-Camelia:2014gga} and show that, unfortunately, this is not compatible with general GUR models.   

In \cite{Amelino-Camelia:2014gga} an ingenious solution to the soccer ball problem was proposed by Amelino-Camelia, who argued that the common formulation of the problem was, in fact, ``a case of mistaken identity''. 
In his proposal, the generalised momentum operators of a given modified commutator model are considered as the generators of `generalised' spatial translations, by definition. 
This requires the unitary operator $\hat{U}({\bf X}) := \exp[(i/\hbar){\bf X}.\hat{{\bf P}}]$ to leave the modified $[\hat{X}^{i},\hat{P}_{j}]$, $[\hat{X}^{i},\hat{X}^{j}]$ and $[\hat{P}_{i},\hat{P}_{j}]$ commutators, as well as the multi-particle Hamiltonian of the model, $\hat{H} := \sum_{I=1}^{N}\hat{\bf{P}}_I^2/(2m_I) + V(\hat{\bf{X}}_1, \hat{\bf{X}}_2, \dots \hat{\bf{X}}_N)$, where the subscript $I$ labels the particle number, invariant. 

Amelino-Camelia's key observation was that, if these invariances hold in a given model, then the corresponding Noether charge for an $N$-particle state is represented by the operator $\hat{{\bf P}}_{\rm Total} := \sum_{I=1}^{N}\hat{{\bf P}_{I}}$, which automatically commutes with the Hamiltonian: $[\hat{{\bf P}}_{\rm Total},\hat{H}] = 0$. 
The usual law of linear momentum addition then holds for multi-particle states but a different nonlinear addition law, derived from the notion of spatial locality, holds for transfers of momentum between individual particles, due to the interactions specified by $\hat{H}$. 
In general, this requires the interaction potential $V(\hat{\bf{X}}_1, \hat{\bf{X}}_2, \dots \hat{\bf{X}}_N)$ to be carefully chosen so that $\sum_{I=1}^{N}\partial V/\partial {\bf X}_{I} = 0$, but this was shown to be possible in a specific example model containing two particles with equal masses, $m_{A} = m_{B} = m$, interacting in a 2-dimensional plane \cite{Amelino-Camelia:2014gga}.  
 
Unfortunately for GUP models, in the example system considered in \cite{Amelino-Camelia:2014gga}, the definition of generalised spatial translation required to maintain the linear addition law for multi-particle states also requires the relation $[\hat{X}^{i},\hat{P}_{i}] = 0$ to hold, for some $i$, for both particles. 
In this case, there is no Heisenberg uncertainty principle, in at least one of the spatial dimensions, let alone a GUP, even though a minimum length-scale $l$ still appears in the model via the position-position commutator, $[\hat{X}_{I}^1,\hat{X}_{I}^2] = il\hat{X}_{I}^1$ .

This illustrates a more general point: it is by no means certain that a particular modified momentum operator, corresponding to a particular modification of the canonical Heisenberg algebra, and, therefore, a particular form of GUR, is compatible with a linear addition law for multi-particle states, derived via the procedure outlined in \cite{Amelino-Camelia:2014gga}. 
In our view, this is not a weakness of Amelino-Camelia's method, which successfully demonstrates that {\it certain classes} of modified commutator models do not, in fact, suffer from a soccer ball problem after all. 
Instead, it is an inherent weakness of models that seek to derive the GUP, EUP or EGUP from modified commutation relations. 
In this respect, the analysis given in \cite{Amelino-Camelia:2014gga} still represents a huge step forward in understanding this problem, and we may apply Amelino-Camelia's procedure to any prospective GUP model based on modified commutators, using it to rule out the ones that give rise to these kinds of inconsistencies in the multi-particle limit. 

In summary, although consistency with Amelino-Camelia's procedure represents a useful criterion for identifying physically viable theories, it is clear that arbitrary deformations of the canonical Heisenberg algebra are not consistent with the existence of a linear momentum addition law for multi-particle states. 
Further work is therefore needed to determine which GUR models truly suffer from a soccer ball problem and which ones do not. 
Though some GUP-type models {\it may} be free from this pathology, it is likely that a great many are still afflicted by it. 
It is therefore clear that the soccer ball problem has not been resolved, in general, for arbitrary GUP, EUP or EGUP models based on modified commutation relations. 

\subsection{Reference frame-dependence of the `minimum' length} \label{Sec.2.4}

In their pioneering and hugely influential work \cite{Kempf:1994su}, Kempf, Mangano and Mann (KMM) gave the first truly rigorous treatment of modified commutator models, showing how they can be derived from the Hilbert space structure of a modified quantum formalism. 
Their key observation was that modified commutators correspond to modified phase space volumes. 
For example, the commutator 
\begin{eqnarray} \label{KMM_mod_comm}
[\hat{X}^i, \hat{P}_j] = i\hbar \delta^{i}{}_{j}(1+\alpha {\bf \hat{P}}^2) \hat{\mathbb{I}} \, ,
\end{eqnarray}
where $\alpha = \alpha_0(m_{\rm Pl}c)^{-2}$ and $\alpha_0$ is a dimensionless constant of order one, which leads to the GUP-type relation 
\begin{eqnarray} \label{KMM_GUP}
\Delta_{\psi}X^i \Delta_{\psi}P_j \geq \frac{\hbar}{2}\delta^{i}{}_{j}(1 + \alpha[(\Delta_{\psi}{\bf P})^2 + \langle\hat{{\bf P}}\rangle_{\psi}^2]) \, ,
\end{eqnarray}
corresponding to a modified normalisation condition and a modified resolution of the identity of the form
\begin{eqnarray} \label{KMM_res_ident}
\langle{\bf P}|{\bf P'}\rangle = (1+\alpha {\bf P}^2)\delta^{3}({\bf P} - {\bf P'}) \, , \quad \int |{\bf P}\rangle\langle{\bf P}| \frac{{\rm d}^3{\rm P}}{(1+\alpha{\bf P}^2)} = \hat{\mathbb{I}} \, .
\end{eqnarray}
Similarly, the commutator 
\begin{eqnarray} \label{KMM_mod_comm*}
[\hat{X}^i, \hat{P}_j] = i\hbar \delta^{i}{}_{j}(1+\eta {\bf \hat{X}}^2) \hat{\mathbb{I}} \, ,
\end{eqnarray}
where $\eta = \eta_0 l_{\rm dS}^{-2}$ and $\eta_0$ is a dimensionless constant of order unity, which leads to the EUP-type relation 
\begin{eqnarray} \label{KMM_GUP*}
\Delta_{\psi}X^i \Delta_{\psi}P_j \geq \frac{\hbar}{2}\delta^{i}{}_{j}(1 + \eta[(\Delta_{\psi}{\bf X})^2 + \langle\hat{{\bf X}}\rangle_{\psi}^2]) \, ,
\end{eqnarray}
corresponding to the modified phase space structure
\begin{eqnarray} \label{KMM_res_ident*}
\langle{\bf X}|{\bf X'}\rangle = (1+\eta {\bf X}^2)\delta^{3}({\bf X} - {\bf X'}) \, , \quad \int |{\bf X}\rangle\langle{\bf X}| \frac{{\rm d}^3{\rm X}}{(1+\eta{\bf X}^2)} = \hat{\mathbb{I}} \, .
\end{eqnarray}

In general, introducing a modified momentum space volume $G({\bf P})^{-1}{\rm d}^3{\rm P}$ ($G({\bf P}) \neq 1$) yields a GUP-type relation, though in this case the position space representation is not well defined, whereas introducing a modified position space volume $G({\bf X})^{-1}{\rm d}^3{\rm X}$ ($G({\bf X}) \neq 1$) yields an EUP-type relation, although the momentum space representation is not well defined. 
For an EGUP-type relation, characterised by the function $G({\bf X},{\bf P}) \neq 1$, neither the position nor momentum space representations are well defined and one must instead introduce a generalised Bargman-Fock representation \cite{Kempf:1996ss}.

To illustrate the problems with these type of constructions, we will focus on the most famous example, proposed in the original KMM paper \cite{Kempf:1994su}, i.e., the GUP-type relation (\ref{KMM_GUP}). 
It is straightforward to show that Eq. (\ref{KMM_GUP}) implies the existence of a `minimum' position uncertainty, $(\Delta_{\psi}X^i)_{\rm min}$, and a corresponding critical value of the momentum uncertainty, $(\Delta_{\psi}P_j)_{\rm crit}$, of the form
\begin{eqnarray} \label{KMM_min_length}
(\Delta_{\psi}X^i)_{\rm min} = \hbar\sqrt{\alpha(1+\alpha \langle\hat{{\bf P}}\rangle_{\psi}^2)} \, , \quad (\Delta_{\psi}P_j)_{\rm crit} = 1/\sqrt{\alpha(1+\alpha \langle\hat{{\bf P}}\rangle_{\psi}^2)} \, .
\end{eqnarray}
The problem with these expressions is that, while the standard deviations on the left-hand sides should be manifestly frame-independent, the quantities on the right are not, since $\langle\hat{{\bf P}}\rangle_{\psi}^2$ is not invariant under Galilean velocity boosts. 

To show this more concretely, let us consider the action of the unitary operator \cite{Lake:2020rwc}
\begin{eqnarray} \label{KMM_p-space_transl}
\tilde{\mathcal{U}}({\bf P'}) |\bf{P}\rangle = \sqrt{\frac{1+\alpha {\bf P}^2}{1+\alpha ({\bf P - P'})^2}} |\bf{P - P'}\rangle \, .
\end{eqnarray}
This generates Galilean velocity boosts, which remain consistent with the modified momentum space volume (\ref{KMM_res_ident}), and reduces to the canonical boost operator $\hat{U}(\bf{p}') |\bf{p}\rangle =  |\bf{p}-\bf{p}'\rangle$ when $\alpha = 0$. 
Its action on the moments of the generalised momentum operator,
\begin{eqnarray} \label{KMM_P}
\hat{P}_{j} = \int P_j |{\bf P}\rangle\langle{\bf P}| \frac{{\rm d}^3{\rm P}}{(1+\alpha{\bf P}^2)} \, ,
\end{eqnarray}
gives
\begin{eqnarray} \label{P_op_transf}
\hat{{\bf P}}^n \mapsto \tilde{\mathcal{U}}({\bf P'}) \hat{{\bf P}}^n \tilde{\mathcal{U}}^{\dagger}({\bf P'}) = (\hat{{\bf P}} + {\bf P'})^n \, ,
\end{eqnarray}
for $n \in \mathbb{N}$, and it is straightforward to demonstrate that this leaves $\Delta_{\psi}P_j$ unchanged. 
The modified commutator (\ref{KMM_GUP}) then transforms as
\begin{eqnarray} \label{KMM_comm_transf}
[\hat{X}^i, \hat{P}_j] \mapsto \tilde{\mathcal{U}}({\bf P'})[\hat{X}^i, \hat{P}_j]\tilde{\mathcal{U}}^{\dagger}({\bf P'}) = i\hbar \delta^{i}{}_{j}(1+\alpha ({\bf \hat{P}}+{\bf P'})^2) \hat{\mathbb{I}} \, ,
\end{eqnarray}
which leads to ${\bf P'}$-dependence of the corresponding uncertainty principle. 
Since $\Delta_{\psi}P_j$ is invariant, it is clear that this must be due to the ${\bf P'}$-dependence of $\Delta_{\psi}X^i$. 

Next, let us denote the boosted position uncertainty as $\Delta_{\psi}X'^i({\bf P'})$, so that $\Delta_{\psi}X'^i(0) \equiv \Delta_{\psi}X^i$, where $ \Delta_{\psi}X^i$ is the position uncertainty given in Eq. (\ref{KMM_GUP}). 
We then have
\begin{eqnarray} \label{}
\Delta_{\psi}X'^i({\bf P'}) \geq \frac{\hbar}{2 \Delta_{\psi}P_j}\delta^{i}{}_{j}(1 + \alpha[(\Delta_{\psi}{\bf P})^2 + (\langle\hat{{\bf P}}\rangle_{\psi} + {\bf P'})^2]) \, . 
\end{eqnarray}
Hence, even if $\psi(\bf{P})$ is symmetric in the original frame of observation, such that $\langle \hat{\bf{P}} \rangle_{\psi} = 0$, the minimum position uncertainty seen by an observer moving with relative velocity ${\bf V'} = {\bf P'}/m$ is
\begin{eqnarray} \label{}
(\Delta_{\psi}X'^i)_{\rm min}({\bf P'}) \simeq \hbar \sqrt{\alpha}\left(1 + \frac{\alpha {\bf P'}^2}{2}\right) \, . 
\end{eqnarray}
For $|{\bf P'}| \ll 1/\sqrt{\alpha} \simeq m_{\rm Pl}c$ the boost-dependent term is, of course, very small, but its presence clearly violates the Galilean boost invariance that emerges from the low velocity limit of Lorentz invariance. 
Its presence is therefore at odds with the founding principles of both special and general relativity, even for one-particle states. 
Analogous reasoning demonstrates the frame-dependence of the `minimum' momentum implied by the EUP-type relation (\ref{KMM_GUP*}), which is no longer invariant under spatial translations.

Though it is possible that boost and/or translation invariance may be broken due to quantum effects on the geometry of spacetime, we note that there is, intrinsically, nothing quantum mechanical about the physical space background of the KMM model. 
The geometry remains classical but its symmetries are unknown, as is the exact form of the classical metric, $g_{ij}(X)$, to which they correspond. 
In Secs. \ref{Sec.2.5}-\ref{Sec.2.6}, we consider the implications of both these points and argue that they lead to further inconsistencies in the modified commutator paradigm. 

\subsection{The background geometry is {\it not} quantum} \label{Sec.2.5}

In the existing literature, there are many references to the `quantum' geometry obtained by introducing modified phase space volumes, as in the KMM model \cite{Kempf:1994su}. 
This motivates a class of so-called nonlocal gravity models that, it is claimed, follow from the `quantum gravity' corrections implied by GURs. 
In this section, we examine the link between modified commutation relations and the proposed nonlocality of the background geometry, and find that this claim does not hold up to scrutiny. 
Instead, we find that the background geometry implied by modified phase space volumes is certainly `classical', in the sense that it does not admit quantum superpositions of states, but that, unlike classical geometries proper, it is not well defined by an appropriate class of symmetries or a metric function, $g_{ij}(X)$. 
These latter considerations are dealt with in detail in Sec. \ref{Sec.2.6}.

Much of the literature on nonlocal gravity models is motivated by the observation that modified momentum space volumes, such as those leading to the GUP-type model proposed in \cite{Kempf:1994su}, can be obtained by acting with an appropriate nonlocal operator on the position space representations of the canonical QM eigenfunctions, 
$\langle{\bf x}|{\bf x}'\rangle = \delta^{3}({\bf x-x'})$ and $\langle{\bf x}|{\bf p}\rangle = (2\pi\hbar)^{-3/2}e^{i{\bf p}.{\bf x}/\hbar}$. 
A simple example is the operator $e^{l^2\Delta}$, where $l$ is a fundamental length scale, usually identified with the Planck length, and $\Delta$ is the Laplacian. 
For convenience, we rewrite this in the spectral representation as
\begin{eqnarray} \label{nonlocal_op}
\hat{\zeta} = e^{-\hat{H}_0\Delta t/\hbar} \, , 
\end{eqnarray}
where $\hat{H}_0 = \hat{p}^2/2m$ is the canonical free particle Hamiltonian and $\Delta t = 2m l^2/\hbar$ is the characteristic time scale associated with $l$ and the particle mass $m$. 
The operator $\hat{\zeta}$ reduces to $e^{l^2\Delta}$ in the wave mechanics picture but we may use Eq. (\ref{nonlocal_op}) to define its action directly on the canonical eigenstates, $|{\bf x}\rangle$ and $|{\bf p}\rangle$, instead of the eigenfunctions, $\langle{\bf x}|{\bf x}'\rangle$ and $\langle{\bf x}|{\bf p}\rangle$. 

This may not seem, at first, like an important distinction, but it is crucial to recognise that $|{\bf x}\rangle$ and $\psi(\bf{x})$ have dimensions of ${\rm (length)^{-3/2}}$ whereas $\langle{\bf x}|{\bf x}'\rangle$ and $|\psi({\bf x})|^2$ have dimensions of ${\rm (length)^{-3}}$. 
Therefore, the position eigenfunction has the dimensions of a {\it probability density}, whereas the position eigenvector has the dimensions of a {\it quantum probability amplitude}. 
This matters because probability densities, including the Dirac delta, are inherently classical in nature, even when they are derived from an underlying quantum mechanical amplitude. 
It is elementary to rewrite any classical probability distribution as the square of a complex distribution, $\rho({\bf x}) \equiv \rho_{\psi}({\bf x}) = |\psi({\bf x})|^2$, but this does not imply that it is quantum mechanical in origin. 
Furthermore, even if it {\it is} quantum mechanical in origin, measurements that depend on $|\psi({\bf x})|^2$ alone destroy all phase information, and are operationally indistinguishable from outcomes that depend only on classical probabilities (i.e., those based on incomplete information about the system) \cite{Ish95}.

Because of this, delocalising the canonical eigen{\it functions}, rather than the eigen{\it states}, maps classical point charges to classical charge-densities of nonzero volume, but does not introduce a genuine quantum state, i.e., a vector in a complex Hilbert space, corresponding to the quantum state of the background geometry. 
Though it is not very flattering to state it in this way, applying the operator $e^{l^2\Delta}$ to $\delta^{3}({\bf x-x'})$ blows up classical point-masses to classical golf balls or grapefruits, but does not achieve much else. 
Furthermore, it is not at all clear whether a classical geometry, in which each zero-dimensional point has somehow been blown up to the size of a three-dimensional grapefruit (or Planck volume), is really a well defined object. 
In the final subsection, Sec. \ref{Sec.2.6}, we will argue that modified phase space volumes cannot be consistently defined in a {\it physical} geometry, but, before that, we consider the action of $\hat{\zeta}$ on $\langle{\bf x}|{\bf x}'\rangle$ and $\langle{\bf x}|{\bf p}\rangle$ in more detail and show, explicitly, that the nonlocal geometry generated by this action is {\it classical}. 

In the usual approach to nonlocal geometry models, the braket $\langle{\bf x}|\hat{\zeta}|{\bf x}'\rangle$ is used to define a set of generalised basis vectors, $|{\bf X}\rangle$ and $|{\bf P}\rangle$, such that
\begin{eqnarray} \label{<X|X'>}
\langle{\bf x}|\hat{\zeta}|{\bf x}'\rangle = e^{\sigma^2\Delta}\langle{\bf x}|{\bf x}'\rangle = \left(\frac{1}{\sqrt{2\pi} \sigma}\right)^3e^{-({\bf x-x'})^2/2\sigma^2} \equiv \langle{\bf X}|{\bf X}'\rangle = \left(\frac{1}{\sqrt{2\pi} \sigma}\right)^3e^{-({\bf X-X'})^2/2\sigma^2} \, , 
\end{eqnarray}
and
\begin{eqnarray} \label{<P|P'>}
\langle{\bf x}|\hat{\zeta}|{\bf p}\rangle = e^{\sigma^2\Delta}\langle{\bf x}|{\bf p}\rangle = \left(\frac{1}{\sqrt{2\pi\hbar}}\right)^3 e^{-{\bf p}^2/2\tilde{\sigma}^2} e^{i{\bf p}.{\bf x}/\hbar} \equiv \langle{\bf X}|{\bf P}\rangle = \left(\frac{1}{\sqrt{2\pi\hbar}}\right)^3 e^{-{\bf P}^2/2\tilde{\sigma}^2} e^{i{\bf P}.{\bf X}/\hbar} \, ,
\end{eqnarray}
where we have rewritten $\sigma \equiv l$ and defined $\tilde{\sigma} \equiv \hbar/\sqrt{2}l$, for later convenience. 
In other words, the nonlocal operator $\hat{\zeta}$ maps the Dirac delta to a finite-volume Gaussian of width $\sigma$ via
\begin{eqnarray} \label{}
\hat{\zeta} : \langle{\bf x}|{\bf x}'\rangle  = \delta^{3}({\bf x}-{\bf x}') \mapsto \langle{\bf X}|{\bf X}'\rangle = \left(\frac{1}{\sqrt{2\pi} \sigma}\right)^3e^{-({\bf X-X'})^2/2\sigma^2} \, , 
\end{eqnarray}
and Eq. (\ref{<P|P'>}) shows its corresponding action on the plane wave $\langle{\bf x}|{\bf p}\rangle = (2\pi\hbar)^{-3/2}e^{i{\bf p}.{\bf x}/\hbar}$. 
Consistency then requires the $|{\bf P}\rangle$ eigenstates to satisfy the modified normalisation condition and modified resolution of the identity
\begin{eqnarray} \label{}
\langle{\bf P} | {\bf P'}\rangle = e^{{\bf P}^2/2\tilde{\sigma}^2}\delta^{3}({\bf P-P'}) \, , \quad \int |{\bf P}\rangle\langle{\bf P}| e^{-{\bf P}^2/2\tilde{\sigma}^2} {\rm d}^3{\rm P} = \hat{\mathbb{I}} \, .
\end{eqnarray}
Expanding $e^{{\bf P}^2/2\tilde{\sigma}^2}$ to first order generates the GUP-type relation of the KMM model \cite{Kempf:1994su}, but the full expressions differ nonperturbatively. 

It is then claimed that the link between GURs and nonlocal gravity is provided by the semi-classical approach \cite{Nicolini:2010dj,Lake:2022hzr} in which the classical curvature of space-time is sourced by the expectation value of the energy-momentum operator for the quantum matter fields, $\langle\hat{T}_{\mu\nu}\rangle_{\psi}$. 
In the weak field limit, the semi-classical field equations reduce to Poisson's equation, with the square of the wave function as a source term, neglecting the subdominant dark energy contribution \cite{Moller:1959bhz,Rosenfeld:1963,Kelvin:2019esx},
\begin{eqnarray} \label{semi-classical-2}
\nabla^2 {\rm \Phi} = 4\pi G m |\psi |^2 \, .
\end{eqnarray}
It is then noted that the zero-width limit of the wave function, $\Delta_{\psi}{\bf X} \rightarrow 0$, yields a delta function source,
\begin{eqnarray} \label{semi-classical-3}
\lim_{\Delta_{\psi}{\bf X} \rightarrow 0}|\psi |^2 = \delta^3({\bf X-X'}) \, .
 \end{eqnarray}
 The standard procedure, therefore, is to substitute the limiting value (\ref{semi-classical-3}) into (\ref{semi-classical-2}) and, interpreting the former as the position eigenfunction of a quantum mechanical wave function, to act on it with the nonlocal operator $e^{\sigma^2\Delta}$. 
On this basis, it is often claimed that GURs imply nonlocal gravity and, furthermore, that the latter arises from `quantum' corrections to the classical geometry. 
There are two main problems with this interpretation. 
First, that in the semi-classical approach on which it is based the geometry is still classical and, second, that $\delta^3({\bf X-X'})$ is a limiting value of $|\psi |^2$ (not $\psi$), which is operationally indistinguishable from a classical finite-density mass distribution. 

We stress, again, that the action of the nonlocal operator on a Dirac delta source term blows up a classical point mass into a classical blob of finite density. 
If the blow up is sufficiently strong, in some sense, the classical matter fluid acquires an effective equation of state which makes it stiff enough to resist gravitational collapse, curing the `singularity problem' \cite{Nicolini:2012eu}, but this has nothing to do with any quantum properties, either of the matter, or of the background geometry. 
The claim that the GURs imply nonlocal gravity is therefore somewhat inaccurate. 
Instead, it is more accurate to claim that classical nonlocal gravity models, such as those defined by Eqs. (\ref{<X|X'>})-(\ref{<P|P'>}), or similar constructions, and modified commutator models, such as the GUP- and EUP-type relations defined by Eqs. (\ref{KMM_res_ident}) and (\ref{KMM_res_ident*}), respectively, stem from the same underlying assumptions, that is, assumed modifications of the classical phase space volumes, over which both classical densities and quantum mechanical amplitudes (wave functions) must be integrated. 

In both cases, the number of degrees of freedom remains the same as in the corresponding local theory and no new quantum degrees of freedom, capable of corresponding to the quantum state of the geometry, are introduced. 
Thus, although the background geometry of modified commutator models is `nonlocal', in some sense, it is {\it not} nonlocal due to quantum effects. 
Furthermore, it is by no means clear whether such `classical nonlocality' is well defined. 
We address this point, in detail, in the following section. 

\subsection{Mathematical inconsistency of modified phase space volumes} \label{Sec.2.6}

In the standard prescription, GURs require modified commutators and modified commutators require modified phase space volumes, yielding a one-to-one correspondence between the two. 
For example, for the KMM GUP-type model, this relation is as follows \cite{Kempf:1994su}:
\begin{eqnarray} \label{KMM}
[\hat{X}^i, \hat{P}_j] = i\hbar \delta^{i}{}_{j}(1+\alpha {\bf \hat{P}}^2) \hat{\mathbb{I}} \iff \hat{P}_{j} = \int P_j |{\bf P}\rangle\langle{\bf P}| \frac{{\rm d}^3{\rm P}}{(1+\alpha{\bf P}^2)} \, .
\end{eqnarray}
 In this section, we explore a number of subtle points that, although {\it implicit} in the construction above, and others like it, have not been explicitly considered in the existing literature. 
 
To begin, we note that models of this form are based on a canonical quantisation procedure that maps classical Poison brackets to commutators, $\left\{x^{i},p_{j}\right\}_{\rm PB} \mapsto \frac{1}{i\hbar}[\hat{X}^{i},\hat{P}_{j}]$, and classical Hamiltonians to quantum Hamiltonians, $H = |{\bf p}|^2/(2m) + V({\bf x}) \mapsto \hat{H} = |\hat{{\bf P}}|^2/(2m) + V(\hat{{\bf X}})$. 
Next, we recall that the phase space of classical mechanics, on which these maps are defined, is a symplectic manifold, and, furthermore, that symplectic geometry is a notoriously `loose' form of geometry. 
Unlike the more familiar Riemannian geometry, which corresponds to our experience of everyday life, symplectic structures do not carry any notion of distance, but volumes can be defined through the introduction of an appropriate symplectic 2-form \cite{Frankel:1997ec,Nakahara:2003nw}. 

Defining a new symplectic 2-form defines a new volume element, but this in no way disturbs the symplectic structure of classical Hamiltonian systems \cite{Frankel:1997ec,Nakahara:2003nw}. 
When the canonical quantisation procedure is applied, this symplectic structure is taken over, unchanged, by the corresponding quantum theory. 
Hence, if the volume element in the classical phase space is, say, $(1+\alpha {\bf P}^2)^{-1}{\rm d}^{3}{\rm x}{\rm d}^{3}{\rm P}$, the phase space volume in the corresponding quantum theory is, simply, $(1+\alpha {\bf P}^2)^{-1}{\rm d}^{3}{\rm P}$. 
It is then assumed that the quantum state vector can be expanded in the `usual' form, except for the modification of the volume element, $| \psi \rangle = \int \psi({\bf P})|{\bf P}\rangle (1+\alpha {\bf P}^2)^{-1}{\rm d}^{3}{\rm P}$, without issue. 
But is this really the case?

To answer this question, we must consider the meanings of the symbols ${\bf P}$ and $\hat{{\bf P}}$, appearing in Eqs. (\ref{KMM}), carefully. 
This is more difficult than it seems, since it is seldom explicitly stated, in the existing GUP literature, what exactly these symbols represent. 
The most probable reason for this is that, of course, everyone already `knows' what they mean: ${\bf P}$ is a momentum space displacement vector and $\hat{{\bf P}}$ is its vector operator counterpart. 
The former may be written as
\begin{eqnarray} \label{P*}
{\bf P} = P_{j} \, {\bf e}^{j}({\rm X}) = P_{X} \, {\bf e}^{X}({\rm X}) + P_{Y} \, {\bf e}^{Y}({\rm X}) + P_{Z} \, {\bf e}^{Z}({\rm X}) \, , 
\end{eqnarray}
where $(X,Y,Z)$ denote {\it global} Cartesian coordinates and $({\bf e}^{X}({\rm X}),{\bf e}^{Y}({\rm X}),{\bf e}^{Z}({\rm X}))$ denote the tangent vectors, to the lines $X = {\rm const.}$, $Y = {\rm const.}$, and $Z = {\rm const.}$, at any point ${\rm X} \in \mathbb{R}^3$ in physical space. 
$\hat{{\bf P}}$ is then constructed in like manner, by replacing the components $P_{j}$ with the operators $\hat{P}_{j}$, defined in Eq. (\ref{KMM}).

The problems with this construction are as follows:
\begin{enumerate}

\item {\it global} Cartesian coordinates only exist in Euclidean space \cite{Frankel:1997ec,Nakahara:2003nw},

\item Euclidean space is a Riemannian geometry, not a symplectic geometry, 

\item it therefore possesses a metric, $g_{ij}({\rm X}) = \langle {\bf e}_{i}({\rm X}),{\bf e}_{j}({\rm X}) \rangle$, which generates both a notion of distance, ${\rm d}L = \sqrt{g_{ij}({\rm X}){\rm d}X^{i}{\rm d}X^{j}}$, and a notion of volume, ${\rm d}V = \sqrt{g}{\rm d}^{3}X$, where $g({\rm X}) = {\rm det}g_{ij}({\rm X})$ is the determinant of the metric, 

\item in the global Cartesians the metric of Euclidean space is $\delta_{ij} = \langle {\bf e}_{i}({\rm X}),{\bf e}_{j}({\rm X}) \rangle$, since ${\bf e}_{i}({\rm X}) = {\bf e}_{i}({\rm X}')$ for all ${\rm X},{\rm X}' \in \mathbb{R}^3$ and $i \in \left\{X,Y,Z\right\}$, giving ${\rm det}\delta_{ij}({\rm X}) = 1$,

\item this gives rise to the distance element $L = \sqrt{X^2 + Y^2 + Z^2}$ and the volume element ${\rm d}V = {\rm d}X{\rm d}Y{\rm d}Z$,

\item the corresponding magnitude of the momentum vector (\ref{P*}) is $P = \sqrt{P_X^2 + P_Y^2 + P_Z^2}$ and the volume element in momentum space is ${\rm d}\tilde{V} = {\rm d}P_X{\rm d}P_Y{\rm d}P_Z$, since the tangent space is isomorphic to the physical Euclidean space \cite{Frankel:1997ec,Nakahara:2003nw},

\item neither of these expressions are flexible.

\end{enumerate}

Unlike volume elements derived from symplectic 2-forms, the Euclidean volume element is fixed by the underlying geometry, as well as the chosen coordinate system. 
The tangent space geometry in which the momentum vector ${\bf P}$ is defined is also fixed, and, in global Cartesians, the lines of constant $X$, $Y$ and $Z$ lie parallel to the trajectories for which $P_X$, $P_Y$ and $P_Z$ are conserved. 
Thus, for the coordinates assumed in the relation $\Delta_{\psi}X^i \Delta_{\psi}P_j \geq \frac{\hbar}{2}\delta^{i}{}_{j}(1 + \alpha[(\Delta_{\psi}{\bf P})^2 + \langle\hat{{\bf P}}\rangle_{\psi}^2])$ (\ref{KMM_GUP}), which is derived from the modified phase space structure (\ref{KMM}), i.e., $i,j \in \left\{X,Y,Z\right\}$, where $(X,Y,Z)$ denote global Cartesians, the volume element of physical space is fixed as ${\rm d}X{\rm d}Y{\rm d}Z$ and the corresponding momentum space volume is ${\rm d}P_X{\rm d}P_Y{\rm d}P_Z$. 

Put simply, if $(X,Y,Z)$ represent global Cartesian coordinates in physical space, $(P_X,P_Y,P_Z)$ represent global Cartesians in the conjugate momentum space. 
This leads to a contradiction, indicating the inconsistency of modified phase space volumes like the one outlined above. 
If we assume that the subscripts $i$ and $j$ in $\Delta_{\psi}X^i$ and $\Delta_{\psi}P_j$ refer to global Cartesian coordinates, then the momentum space integration measure is simply ${\rm d}^{3}P$. 
Conversely, if we abandon this assumption and {\it define} ${\bf P}^2$ such that ${\bf P}^2 \equiv P^2 = \sum_{j=1}^{3}P_j^2$, without specifying the coordinates $(X^{1},X^{2},X^{3})$, we are faced with an even bigger problem: in this case, we cannot make any physical predictions at all!

As stated in the Introduction, we believe that this constitutes the most serious criticism of the modified commutator / modified phase space paradigm yet formulated in the literature. 
It has immediate real world implications since, to the best of our knowledge, existing experimental bounds on the GUP parameter $\alpha$ have all been obtained by adopting two contradictory assumptions: (1) that the GUP arises from a modified commutation relation, and (2) that the modified commutation relation holds for the uncertainties $\Delta_{\psi}X^i$ and $\Delta_{\psi}P_j$ where $i,j \in \left\{X,Y,Z\right\}$ refer to global Cartesian coordinates \cite{Pikovski:2011zk,Bosso:2016ycv,Kumar:2017cka,Girdhar:2020kfl,Cui:2021gsf,Sen:2021oqj}. 

In Sec. \ref{Sec.4}, we show how to derive the GUP, EUP and EGUP for Cartesian uncertainties, without introducing modified commutation relations or phase space volumes. 
The new model avoids the inconsistencies inherent in these models, in an almost trivial way, but is found to have exceedingly nontrivial implications for the quantisation of the background geometry. 
Before that, in Sec. \ref{Sec.3}, we consider whether the problems outlined here might instead be considered as `features', rather than `bugs', of modified commutator models. 

\section{Bug or feature?} \label{Sec.3}

It could be argued that a number of the problems discussed above represent features, rather than bugs, of the modified commutator paradigm. 
For example, it is widely accepted that some kind of breakdown of the EP and / or Lorentz invariance must occur due to quantum gravitational effects. 
Therefore, it may be regarded as unsurprising that minimum length models based on modified commutators generate both as a matter of course. 
From this viewpoint, such features acquire the status of `smoking guns', i.e., predictions of new physics in the low-energy regime that any self-consistent model of high-energy quantum gravity must successfully reproduce.

These are strong claims, and, here, we argue that such claims require strong motivations, which are currently lacking in existing models. 
Furthermore, it is necessary to prove beyond doubt that such modifications of the canonical formalism do not give rise to internal inconsistencies, even within their domain of applicability. 

As an analogy, there are an infinite number of ways to break the Poincar{\' e} group symmetries of Minkowski space, but only one of these gives rise to a self-consistent limit in the non-relativistic regime, i.e., the Galilean group of Euclidean space, with time as a parameter. 
By contrast, arbitrary violations of Poincar{\' e} symmetry do not give rise to well-defined geometries of any kind. 
Similarly, the breaking of a given symmetry, or equivalence, in a physical theory, is not the same as introducing a well-motivated generalisation of the original model. 
Since there are an infinite number of ways to break anything, why should this or that violation be preferred? 
Is the resulting model self-consistent? 
Below, we briefly consider the `bug or feature' argument for each of the problems raised in Secs. \ref{Sec.2.1}-\ref{Sec.2.6}.

\begin{enumerate}

\item Violation of the EP: That the EP must break down due to quantum gravity effects is certain, but this is so for a very simple reason - the standard EP, as formulated in general relativity, is an inherently {\it classical} concept. 
It concerns the equivalence between classical gravity and classical accelerated frames of reference, which, by definition, excludes the concept of quantum superposition. In the quantum gravitational regime, we expect classical geometries to be replaced by quantum superpositions of geometries, and classical accelerated frames of reference to be replaced by quantum superpositions thereof, but is this the same as simply {\it breaking} the equivalence between the existing classical concepts? 
While it is possible, and even likely, that an appropriate course-graining over such superpositions could give rise to discrepancies with the existing classical theory, it is by no means clear what form these ought to take. 
In the absence of any indication from a specific model in the UV sector, we argue that the prediction of mass-dependent accelerations, in a classical background `geometry', without a well-defined symmetry group, should be treated with extreme caution.
\footnote{We recall that modifications of the canonical Heisenberg algebra imply the violation of Galilean symmetries.}
Nonetheless, it must be admitted that this remains a theoretical possibility, which cannot be excluded {\it a priori}, at the present time.

\item Violation of Lorentz invariance: Similar arguments can be made regarding the violation of Lorentz invariance in the relativistic limit. 
We expect the quantisation of the geometric background to introduce space-time superpositions, generalising the classical concept of a sharp space-time point and, with it, the concept of a sharply-defined classical frame of reference, whether accelerated or inertial, to include superpositions. 
But this is not the same as simply breaking Lorentz symmetry in a still-classical background, especially when it is not clear whether the broken symmetry group corresponds to a well-defined geometry or not. 
In general, deforming an algebra of group generators does not yield a well-defined group, and, hence, the resulting operators do not correspond to a well-defined geometry. 
It is not analogous to the shift from relativistic space-time to non-relativistic space-plus-time, but a far more ambiguous procedure. 
Nonetheless, it must again be admitted that an appropriate course-graining over superpositions in the low-energy limit of quantum gravity may, in fact, violate Lorentz invariance in a way predicted by some classes of modified commutator models.

\item The soccer ball problem: It must also be admitted that the partial resolution of the soccer ball problem, presented in \cite{Amelino-Camelia:2014gga}, provides a potential way out of the problems above. 
In this approach, the issue of the background geometry is moot, since the generalised momentum operators of the modified commutation relation are required to generate modified (non-Galilean) symmetries of the Hamiltonian. 
They may then be regarded as the symmetry generators of a modified (non-Euclidean) geometry, by definition. 
However, this places severe constraints on the form these generators may take, and, as shown in Sec. \ref{Sec.2.3}, not all such modifications are compatible with the existence of the GURs suggested by quantum gravity thought experiments. 
Therefore, at present, it is not clear whether this problem should be regarded as a bug, or a feature, of modified commutator models.

\item The reference frame dependence of the minimum length: The situation is different in the case of the reference frame dependence of the minimum length (momentum) scale, predicted by the GUP (EUP). 
This clearly represents a mathematical inconsistency of modified commutator models, since a mere shift in the coordinate origin, or Galilean velocity boost $v \rightarrow v' \ll c$, radically alters the values of the position and / or momentum uncertainties of the quantum wave packet. 
The standard measure of the statistical variance of a random variable, $(\Delta O)^2 = \braket{O^2} - \braket{O}^2$, is constructed to be manifestly invariant under such transformations. 
If this invariance no longer holds in modified commutator models, it is unclear how to measure the spread of the wave function in a coordinate-independent way. 
This undoubtably counts as a `bug'.

\item The background geometry is not quantum: We may also ask if the {\it classically} nonlocal background geometry of the KMM model \cite{Kempf:1994su}, and others like it, could emerge from a more fundamental underlying quantum theory? 
Theoretically, the answer is again `yes', but the same could be said of any (totally arbitrary) modification of the canonical quantum formalism. 
The most important point here is that, at present, no concrete proposal for such a model has been presented in the GUP literature \cite{Tawfik:2014zca,Tawfik:2015rva,Hossenfelder:2012jw}. 
It had been hoped that the nonlocal operator, Eq. (\ref{nonlocal_op}), provided such a link, between the coarse-grained classical structure and an underlying probabilistic (i.e., quantum) model, but this assumption was debunked in Sec. \ref{Sec.2.5}. 
The confusion in the literature stemmed from a confusion between classical probability densities and quantum probability {\it amplitudes}. 
The status of this problem, therefore, resembles that of problems 1-2. 
Theoretically, the classically nonlocal geometry produced by the action of Eq. (\ref{nonlocal_op}) may emerge from an appropriate coarse-graining over quantum probability amplitudes, but no concrete mathematical structure, able to reproduce this, has been discovered.

\item The mathematical inconsistency of modified phase space volumes: In the Introduction, we claimed that this, deceptively simple argument, represents the most serious objection yet raised to the modified commutator paradigm. 
In short, it is certainly not a feature, but a bug, since it represents a serious mathematical inconsistency of such models. 
In light of this, the previous five points raised above may be considered moot, since only one inconsistency is required to render a physical theory untenable. 
It is therefore of the utmost importance to establish whether or not this objection can be circumvented in some way.

\end{enumerate}

\section{Deriving the GUP, EUP and EGUP without modified commutators} \label{Sec.4}

It is straightforward to show that the GUP and EUP can be obtained, at least approximately, from far more `natural' looking expressions, in which the variances of independent random variables add linearly. 
For example, let us consider the simplest scenario in which the back reaction of the wave function on the geometry is considered negligible, so that the latter undergoes quantum fluctuations which are independent of $\psi({\bf x})$. 

In this case, $(\Delta x_{\psi})^2$ denotes the variance of the canonical probability density $|\psi({\bf x})|^2$, where ${\bf x} \in \mathbb{R}^3$ are the possible measured values of the particle's position in classical three-dimensional space. 
These are the canonical quantum degrees of freedom. 
In order to describe quantum fluctuations of the background we must introduce new degrees of freedom which are capable of describing {\it superpositions of geometries}, as expected in a viable theory of quantum gravity \cite{Marletto:2017pjr,Lake:2018zeg}. 

The additional fluctuations in the measured position of the particle, due to quantum fluctuations of the geometry in which it `lives', may be described by an additional variance, $(\Delta_g x')^2$. 
This denotes the variance of the non-canonical probability density $|g({\bf x'-x})|^2$, where $|{\bf x'-x}|$ quantifies the size of the fluctuation, i.e., the degree to which the measured position of the particle is perturbed by the quantum nature of the background. 
(For simplicity, we may imagine $|g({\bf x'-x})|^2$ as a three-dimensional Planck-width Gaussian distribution.) 
This corresponds to a new composite wave function, $\Psi({\bf x},{\bf x}')$, which describes the propagation of a quantum particle in a quantum background, rather than a fixed classical geometry: 
\begin{eqnarray} \label{Psi(x,x')}
\Psi({\bf x},{\bf x}') = \psi({\bf x})g({\bf x'-x}) \, . 
\end{eqnarray}
The possible measured positions of the particle are then given by the values of ${\bf x'} \in \mathbb{R}^3$, rather than ${\bf x}$, and $g({\bf x'-x})$ may be interpreted as the quantum probability amplitude for the coherent transition ${\bf x} \mapsto {\bf x}'$ in a smeared superposition of geometries \cite{Lake:2018zeg,Lake:2019nmn}. 
From here on, we refer to $g$ as the `smearing function'. 
The total variance for a position measurement in the smeared space is then, simply
\begin{eqnarray} \label{Linear_Var_add_X}
(\Delta_{\Psi} x'^{i})^2 = (\Delta_{\psi} x'^{i})^2 + (\Delta_{g} x'^{i})^2 \, . 
\end{eqnarray}
Setting $\Delta x_{g}' \simeq l_{\rm Pl}$, taking the square root of Eq. (\ref{Linear_Var_add_X}) and Taylor expanding to first order then yields the GUP. 
However, in this case, the relevant uncertainties are well defined, as the standard deviations of probability densities, unlike the heuristic uncertainties given in (\ref{GUP-1}).  

A similar construction in momentum space, 
\begin{eqnarray} \label{Psi(p,p')}
\tilde{\Psi}({\bf p},{\bf p}') = \tilde{\psi}({\bf p})\tilde{g}({\bf p'-p}) \, ,
\end{eqnarray}
yields 
\begin{eqnarray} \label{Linear_Var_add_P}
(\Delta_{\Psi} p'_{j})^2 = (\Delta_{\psi} p'_{j})^2 + (\Delta_{g} p'_{j})^2 \, . 
\end{eqnarray}
Setting $\Delta p_{g}' \simeq m_{\rm dS}c$, taking the square root of Eq. (\ref{Linear_Var_add_P}) and Taylor expanding to first order then yields the EUP. 
Again, the relevant uncertainties are well defined, as the standard deviations of the probability densities $|\tilde{\psi}({\bf p})|^2$ and $|\tilde{g}({\bf p'-p})|^2$, unlike the heuristic uncertainties in (\ref{EUP-1}). 

But what, exactly, are the functions $\tilde{\psi}({\bf p})$ and $\tilde{g}({\bf p'-p})$? 
How are they related to $\psi({\bf x})$ and $g({\bf x'-x})$? 
In canonical QM, $\tilde{\psi}({\bf p})$ is the $\hbar$-scaled Fourier transform of $\psi({\bf x})$ and, to emphasize this point, we rewrite it with an appropriate subscript as
\begin{eqnarray} \label{tilde_psi}
\tilde{\psi}({\bf p}) \equiv \tilde{\psi}_{\hbar}({\bf p}) = \left(\frac{1}{\sqrt{2\pi\hbar}}\right)^3 \int \psi({\bf x}) e^{-\frac{i}{\hbar}{\bf p}.{\bf x}}{\rm d}^3{\rm x} \, . 
\end{eqnarray}
The canonical de Broglie relation ${\bf p} = \hbar {\bf k}$ ensures that the exponent is independent of Planck's constant, but $\hbar$ necessarily appears in this expression through the normalisation constant $\sqrt{2\pi\hbar}^{-3}$. 
This is because $|\psi\rangle$ represents the state of a canonical quantum particle and $\hbar$ sets the (action) scale at which quantum effects become significant in canonical quantum matter \cite{Rae}. 

However, were we to assume, likewise, that $\tilde{g}({\bf p'-p})$ is given by the canonical $\hbar$-scaled Fourier transform of $g({\bf x'-x})$, we would obtain the expression (\ref{Linear_Var_add_P}) with $\Delta_{g} p' \simeq m_{\rm Pl}c$ not 
$\Delta_{g}p' \simeq m_{\rm dS}c$! 
This leads to an EUP-type expression with a `minimum' momentum equal to the Planck momentum and, hence, a minimum energy equal to the Planck energy, a minimum energy density equal to the Planck density, etc., which is clearly at odds with empirical data. 
Therefore, we must instead assume a decomposition of the form
\begin{eqnarray} \label{tilde_g}
\tilde{g}({\bf p}' - {\bf p}) \equiv \tilde{g}_{\beta}({\bf p}'-{\bf p}) = \left(\frac{1}{\sqrt{2\pi\beta}}\right)^3 \int g({\bf x}'-{\bf x}) e^{-\frac{i}{\beta}({\bf p}'-{\bf p}).({\bf x}'-{\bf x})}{\rm d}^3{\rm x}' \, ,
\end{eqnarray}
with $\beta \neq \hbar$. 
It is straightforward to show that, in order to recover both the GUP (\ref{GUP-1}) from Eq. (\ref{Linear_Var_add_X}) {\it and} the EUP (\ref{EUP-1}) from Eq. (\ref{Linear_Var_add_P}), we must set
\begin{eqnarray} \label{smear_scales}
\Delta_g x'^{i} = \sigma_g := \sqrt{2\alpha_0}l_{\rm Pl} \, , \quad \Delta_g p_{j}' = \tilde{\sigma}_g := \sqrt{6\eta_0}m_{\rm dS}c \, , \quad \forall i,j \, ,
\end{eqnarray}
together with 
\begin{eqnarray} \label{beta-1}
\beta := 2 \sigma_g\tilde{\sigma}_g \, . 
\end{eqnarray}
For $\alpha_0$, $\eta_0 \sim \mathcal{O}(1)$, this gives
\begin{eqnarray} \label{beta-2}
\beta = 2\sqrt{\frac{\rho_{\Lambda}}{\rho_{\rm Pl}}}\hbar \simeq 10^{-61} \times \hbar \, , 
\end{eqnarray}
where $\rho_{\Lambda} = \Lambda c^2/(8\pi G) \simeq 10^{-30}$ ${\rm g} \, . \, {\rm cm}^{-3}$ is the dark energy density \cite{Perlmutter1999,Reiss1998} and $\rho_{\rm Pl} = c^5/(\hbar G^2) \simeq 10^{93}$ ${\rm g} \, . \, {\rm cm}^{-3}$ is the Planck density. 
Taken together, Eqs. (\ref{tilde_psi}) and (\ref{tilde_g}) are equivalent to imposing the modified de Broglie relation, 
\begin{eqnarray} \label{mod_dB}
{\bf p} = \hbar{\bf k} + \beta({\bf k}' - {\bf k}) \, , 
\end{eqnarray}
where the non-canonical term can be understood, heuristically, as the additional momentum `kick' imparted to the canonical quantum state, by a fluctuation of the background \cite{Lake:2018zeg,Lake:2019nmn}. 

Equations (\ref{Linear_Var_add_X}) and (\ref{Linear_Var_add_P}) can also be recovered, with the appropriate minimum values (\ref{smear_scales}), from the canonical-type braket constructions
\begin{eqnarray} \label{X-braket}
(\Delta_{\Psi} X^{i})^2 = \langle \Psi | (\hat{X}^{i})^2 | \Psi \rangle - \langle \Psi | \hat{X}^{i} | \Psi \rangle^2 \, ,  
\end{eqnarray}
\begin{eqnarray} \label{P-braket}
(\Delta_{\Psi} P_{j})^2 = \langle \Psi | (\hat{P}_{j})^2 | \Psi \rangle - \langle \Psi | \hat{P}_{j} | \Psi \rangle^2 \, , 
\end{eqnarray}
where we have relabelled $\Delta_{\Psi} x'^{i} \equiv \Delta_{\Psi} X^{i}$ and $\Delta_{\Psi} p'_{j} \equiv \Delta_{\Psi} P_{j}$, for the sake of notational convenience. 
The appropriate generalised operators, $\hat{X}^{i}$ and $\hat{P}_{j}$, representing position and momentum measurements in the smeared superposition of geometries, are given by
\begin{eqnarray} \label{X}
\hat{X}^{i} := \int x'^{i} |{\bf x},{\bf x}'\rangle \langle {\bf x},{\bf x}'|{\rm d}^3{\rm x}{\rm d}^3{\rm x}' \, ,  
\end{eqnarray}
\begin{eqnarray} \label{P}
\hat{P}_{j} := \int p'_{j} |{\bf p}\,{\bf p}'\rangle \langle {\bf p}\,{\bf p}'|{\rm d}^3{\rm p}{\rm d}^3{\rm p}' \, ,  
\end{eqnarray}
where $|{\bf x},{\bf x}'\rangle := |{\bf x}\rangle \otimes |{\bf x}'\rangle$ and the basis $|{\bf p}\,{\bf p}'\rangle$ is entangled, 
\begin{eqnarray} \label{entangled_basis}
|{\bf p}\,{\bf p}'\rangle := \left(\frac{1}{2\pi\sqrt{\hbar\beta}}\right)^{3} \int\int |{\bf x},{\bf x}'\rangle e^{-\frac{i}{\hbar}{\bf p}.{\bf x}} e^{-\frac{i}{\beta}({\bf p}'-{\bf p}).({\bf x}'-{\bf x})}{\rm d}^3{\rm x}{\rm d}^3{\rm x}' \, .
\end{eqnarray}
(We emphasize this by not writing a comma between ${\bf p}$ and ${\bf p}'$.) 

However, the position and momentum space bases may be symmetrized, such that $|{\bf x},{\bf x}'\rangle \mapsto |{\bf x},{\bf x}' - {\bf x}\rangle$ and $|{\bf p}\,{\bf p}'\rangle \mapsto |{\bf p},{\bf p}' - {\bf p}\rangle$, by means of an appropriate unitary transformation \cite{Lake:2019nmn}. 
Formally, this is analogous to the unitary transformation defined in \cite{Giacomini:2017zju}, which is intended to represent a switch between quantum reference frames (QRFs), but with the substitution $\beta \leftrightarrow \hbar$. 
This implies that, in the quantum mechanical `smeared space' defined by our model, each `smeared point' may be considered as a QRF, whose quantum uncertainties are controlled by the quantum of action for geometry, 
$\beta$, rather than that for canonical quantum matter, $\hbar$ \cite{Lake:2020chb,Lake:2021beh,Lake:2021gbu}.

At first glance, the introduction of a second quantisation scale for geometry appears to contradict a rather large body of existing literature which claims that Planck's constant is unique. 
A closer look at this literature, however (for example, see \cite{Sahoo2004,Deser:2022lmi} and references therein), shows that only quantisation schemes of the form ${\bf p} = \hbar{\bf k}$, ${\bf p}' = \hbar'{\bf k}'$, with $\hbar' \neq \hbar$, where ${\bf p}$ and ${\bf p}'$ refer to the momenta of different species of material particles, are ruled out by existing no-go theorems. 
The crucial mathematical difference between these models and the modified de Broglie relation (\ref{mod_dB}) is the presence of relative variables, ${\bf k}' - {\bf k}$, in the latter. 
Physically, this is directly related to our treatment of the composite matter-geometry system as a QRF, in which the relative variable ${\bf x}' - {\bf x}$ describe quantum fluctuations of the smeared spatial points \cite{Lake:2018zeg,Lake:2019nmn}. 
This physical picture leads directly to a well-defined EGUP since Eqs. (\ref{Linear_Var_add_X}), (\ref{Linear_Var_add_P}), (\ref{smear_scales}) and (\ref{X-braket})-(\ref{P-braket}) can also be combined, directly, to give \cite{Lake:2019nmn}: 
\begin{eqnarray} \label{EGUP-2}
\Delta_{\Psi} X^i\Delta_{\Psi} P_j \gtrsim \frac{\hbar}{2} \delta^{i}{}_{j} \left[1 + \alpha_0 \frac{2G}{\hbar c^3}(\Delta_{\Psi} P_j)^2 + 2\eta_0\Lambda (\Delta_{\Psi} X^i)^2\right] \, .
\end{eqnarray} 

Crucially, it is straightforward to show that the smeared space position and momentum operators (\ref{X})-(\ref{P}) obey the following commutation relations:  
\begin{eqnarray} \label{rescaled_Heisenberg-A}
[\hat{X}^{i},\hat{P}_{j}] = i(\hbar + \beta) \delta^{i}{}_{j} \, \hat{\mathbb{I}} \, , 
\end{eqnarray}
\begin{eqnarray} \label{rescaled_Heisenberg-B}
[\hat{X}^{i},\hat{X}^{j}] = 0 \, , \quad [\hat{P}_{i},\hat{P}_{j}] = 0 \, . 
\end{eqnarray}
Since these are just a rescaled version of the canonical Heisenberg algebra, with $\hbar \mapsto \hbar + \beta$, we obtain the `expected' quantum gravity phenomenology, i.e., the GUP, EUP and EGUP (\ref{EGUP-2}), without introducing any of the pathologies associated with standard modified commutator models \cite{Lake:2020rwc,LakeUkraine2019}. 
The physical interpretation of this rescaled algebra is subtle and the interested reader is referred to the more in-depth and complete works \cite{Lake:2019nmn,Lake:2020rwc} for further details.

The price we have to pay for this neat solution is the introduction of a second quantisation constant for geometry, $\beta \ll \hbar$, which is directly related to the dark energy scale. 
With this in mind, we note that the product of the position and momentum uncertainties in smeared space, $\Delta_{\Psi} X^i\Delta_{\Psi} P_j$ in the EGUP (\ref{EGUP-2}), is minimized when (see \cite{Lake:2018zeg} for details): 
\begin{eqnarray} \label{opt-1}
\Delta_{\psi} x'^{i} = \sqrt{\frac{\hbar}{2}\frac{\Delta_g x'^{i}}{\Delta_g p'_{i}}} \, , \quad \Delta_{\psi} p'_{j} = \sqrt{\frac{\hbar}{2}\frac{\Delta_g p'_{j}}{\Delta_g x'^{j}}} \, .
\end{eqnarray}
For the minimum values given by Eqs. (\ref{smear_scales}) this yields
\begin{eqnarray} \label{opt-2}
\Delta_{\Psi} X \simeq l_{\Lambda} := \sqrt{l_{\rm Pl}l_{\rm dS}} \simeq 0.1 \, {\rm mm} \, , \quad
c\Delta_{\Psi} P \simeq m_{\Lambda}c^2 := \sqrt{m_{\rm Pl}m_{\rm dS}}c^2 \simeq 10^{-3} \, {\rm eV} \, ,
\end{eqnarray}
where we have neglected to label dimensional indices. 
The corresponding energy density is
\begin{eqnarray} \label{energy_dens}
\mathcal{E}_{\Psi} \simeq \frac{c\Delta_{\Psi} P}{(\Delta_{\Psi} X)^3} \simeq \frac{m_{\Lambda}c}{l_{\Lambda} ^3} \simeq \frac{\Lambda c^4}{G} \simeq \rho_{\Lambda}c^2 \, ,
\end{eqnarray}
so that any field which minimizes the smeared-space uncertainty relations must, necessarily, possess an energy density comparable to the present day dark energy density, $\rho_{\Lambda} \simeq 10^{-30}$ ${\rm g} \, . \, {\rm cm}^{-3}$. 
In this scenario, the immense difference between the matter and geometry quantisation scales may be regarded as `fundamental', while the immense difference between the Planck density and the observed vacuum density is an emergent phenomenon, stemming, ultimately, from the quantum properties of space-time \cite{Lake:2018zeg,Lake:2019nmn,Lake:2020chb}.   

\section{Discussion} \label{Sec.5}

In the first part of this paper we outlined six major pathologies that afflict models of generalised uncertainty relations (GURs) based on modified commutation relations. 
The first two of these, namely, violation of the equivalence principle and violation of Lorentz invariance in the relativistic limit, have been addressed at length in the existing literature \cite{Tawfik:2014zca,Hossenfelder:2012jw,Hossenfelder:2014ifa} and we summarised them only briefly. 
The third, the so-called soccer ball problem for multi-particle states, has also been considered in detail and a would-be solution was proposed in \cite{Amelino-Camelia:2014gga}. 
Though ingenious, and valid within its domain of applicability, we showed that the solution put forward in \cite{Amelino-Camelia:2014gga} does not apply, in general, to arbitrary GUR models based on modifications of the canonical Heisenberg algebra. 
The fourth and fifth problems, the reference frame-dependence of the would-be `minimum' length and the inherently classical nature of the background geometry in modified commutator models, were considered previously in \cite{Lake:2020rwc}, but have not, to the best of our knowledge, been addressed elsewhere. 
Finally, we argued that there is, in fact, a sixth problem that appears in modified commutator models, which, remarkably, has not been considered at all the existing literature. 

This problem is nothing less than the mathematical inconsistency of the modified phase space volumes from which modified commutators, and hence GURs, are usually derived. 
The essence of the problem is that the position-space coordinates $X^{i}$, corresponding to the quantum uncertainty $\Delta X^{i}$, are assumed to represent global Cartesians, $X^{i} \in \left\{X,Y,Z\right\}$. 
The associated distance and volume measures in real space are then given by $L = \sqrt{X^2 + Y^2 + Z^2}$ and $V = {\rm d}X{\rm d}Y{\rm d}Z$, respectively. 
This immediately rules out the existence of modified $X$-space volumes and, hence, modified algebras leading to EUP-type uncertainty relations. 
Likewise, if $X^{i} \in \left\{X,Y,Z\right\}$ form a global Cartesian coordinate system in real space then the conjugate momenta $P_{j} \in \left\{P_X,P_Y,P_Z\right\}$, corresponding to the quantum uncertainty $\Delta P_{j}$, must form a global Cartesian coordinate system in momentum space. 
The associated distance and volume measures are $P = \sqrt{P_X^2 + P_Y^2 + P_Z^2}$ and $\tilde{V} = {\rm d}P_X{\rm d}P_Y{\rm d}P_Z$. 
This immediately rules out the existence of modified $P$-space volumes and, hence, modified algebras leading to GUP-type uncertainty relations. 
To make matters worse, the usual approach in the literature is to assume the validity of the standard $X$- and $P$-space distance measures while simultaneously adopting modified volume forms. 
This procedure is mathematically inconsistent. 

If, instead, we choose to abandon the assumption that $X^{i}$ and $P_{j}$ label global Cartesians in position and momentum space, respectively, we are faced with the following very difficult question: what, exactly, do they represent? 
This question is exceedingly difficult because, unless it can be answered concretely, abstract mathematical expressions of the form $\Delta X^{i}\Delta P_{j} \geq (\hbar/2)\delta^{i}{}_{j}G(\hat{{\bf X}},\hat{{\bf P}})$ cannot be used to make {\it any} valid physical predictions. 
We believe that this deceptively simple observation represents the most serious objection to the modified commutator paradigm yet raised in the literature and that, taken together, the six pathologies described herein ought to signal the `death knell' of modified commutator models. 
Though some have been discussed only recently, all six were inherent in such models from their conception nearly three decades ago, and appear no closer to being solved today than they were in the mid-1990s. 
Indeed, substantial evidence now suggests that at least some of these pathologies {\it cannot} be consistently resolved. 

We believe that this, accumulated evidence, should strongly motivate the GUR research community to seek alternative mathematical structures which are capable of generating the same phenomenology, without the inconsistencies, ambiguities, and headaches associated with modified commutation relations. 
To this end we outlined one such formalsim, originally proposed in a series of works coauthored by one of us \cite{Lake:2018zeg,LakeUkraine2019,Lake:2019nmn,Lake:2021beh,Lake:2020chb,Lake:2021gbu,Lake:2020rwc,Lake:2022hzr,Lake-Frontiers-1}, in the second part of this paper. 
Whether, ultimately, this model has anything to do with physical reality, or not, is perhaps less important than what is demonstrates: that GUP, EUP and EGUP phenomenology can be obtained {\it without} assuming modified commutation relations of a non-Heisenberg type. 
This demonstrates, by means of an explicit example, the logical independence of GURs and modified algebras. 
The latter certainly do imply the former (though not, as we have seen, in a self-consistent formulation) whereas the former do not, in fact, {\it require} them. 
This is a common misconception in the existing literature, in which these two distinct mathematical structures are often conflated. 
There is no one-to-one correspondence between GURs and modified commutation relations and the two are {\it not} logically equivalent. 
This opens up an intriguing and exciting possibility, namely, that other mathematical structures, not yet discovered, are also capable of generating GURs without modified commutators. 
Potentially, these may tell us a great deal about the structure of low-energy quantum gravity and, hence, about the possible structure of a unified theory. 
We implore the phenomenological research community to search for them, earnestly, and to explore their implications as thoroughly as they have explored the implications of modified commutation relations, over the past quarter of a century.



\section*{Acknowledgments}

ML acknowledges the Department of Physics and Materials Science, Faculty of Science, Chiang Mai University, for providing research facilities, and the Natural Science Foundation of Guangdong Province, grant no. 008120251030. AW would like to acknowledge partial support from the Center of Excellence in Quantum Technology, Faculty of Engineering, Chiang Mai University and the NSRF via the Program Management Unit for Human Resources \& Institutional Development, Research and Innovation [grant number B05F640218], National Higher Education Science Research and Innovation Policy Council. 



\begin{thebibliography}{99}

\bibitem{Maggiore:1993rv} 
   M.~Maggiore,
   {\it A Generalized uncertainty principle in quantum gravity},
   Phys.\ Lett.\ B {\bf 304}, 65 (1993).

\bibitem{Adler:1999bu} 
   R.~J.~Adler and D.~I.~Santiago, 
   {\it On gravity and the uncertainty principle},
   Mod. Phys. Lett. A \textbf{14}, 1371 (1999).

\bibitem{Scardigli:1999jh} 
   F.~Scardigli, 
   {\it Generalized uncertainty principle in quantum gravity from micro - black hole Gedanken experiment},
   Phys.\ Lett.\ B \textbf{1999}, {\it 452}, 39.
      
\bibitem{Bolen:2004sq}
   B.~Bolen and M.~Cavaglia,  
   {\it (Anti-)de Sitter black hole thermodynamics and the generalized uncertainty principle},
   Gen. Rel. Grav. \textbf{37}, 1255-1262 (2005).
   
\bibitem{Park:2007az} 
   M.~I.~Park, 
   {\it The Generalized Uncertainty Principle in (A)dS Space and the Modification of Hawking Temperature from the Minimal Length},
   Phys. Lett. B \textbf{659}, 698-702 (2008).

\bibitem{Bambi:2007ty} 
   C.~Bambi, F.~R.~Urban,   
   {\it Natural extension of the Generalised Uncertainty Principle},
   Class. Quant. Grav. \textbf{25}, 095006 (2008).
   
\bibitem{Betoule:2014frx} 
   M.~Betoule, M. {\it et al.} [SDSS Collaboration]. 
   {\it Improved cosmological constraints from a joint analysis of the SDSS-II and SNLS supernova samples},
   Astron. Astrophys. \textbf{568}, A22 (2014).
   
\bibitem{Aghanim:2018eyx}
   N.~Aghanim {\it et al.} [Planck Collaboration].
   {\it Planck 2018 results. VI. Cosmological parameters}, 
   Astron. Astrophys. \textbf{641}, A6 (2020)
   [erratum: Astron. Astrophys. \textbf{652}, C4 (2021)].
   
\bibitem{Tawfik:2014zca} 
   A.~N.~Tawfik and A.~M.~Diab, 
   {\it Generalized Uncertainty Principle: Approaches and Applications},
   Int. J. Mod. Phys. D \textbf{23}, no.12, 1430025 (2014).

\bibitem{Tawfik:2015rva}
   A.~N.~Tawfik and A.~M.~Diab,
   {\it Review on Generalized Uncertainty Principle},
   Rept. Prog. Phys. \textbf{78}, 126001 (2015).
   
\bibitem{Robertson:1929zz} 
   H.~P.~Robertson, 
   {\it The Uncertainty Principle},
   Phys.\ Rev.\  {\bf 34}, 163 (1929).
 
\bibitem{Schrodinger:1930ty} 
   E.~Schr{\" o}dinger,
   {\it About Heisenberg uncertainty relation},
   Bulg.\ J.\ Phys.\  {\bf 26}, 193 (1999)
   [Sitzungsber.\ Preuss.\ Akad.\ Wiss.\ Berlin (Math.\ Phys.\ ) {\bf 19}, 296 (1930)].
   
\bibitem{Hossenfelder:2012jw} 
   S.~Hossenfelder,  
   {\it Minimal Length Scale Scenarios for Quantum Gravity},
   Living Rev. Rel. \textbf{16}, 2 (2013).

\bibitem{Hossenfelder:2014ifa} 
   S.~Hossenfelder, 
   {\it The Soccer-Ball Problem},
   SIGMA \textbf{10}, 074 (2014).
   
 
\bibitem{Lake:2020rwc}
   M.~J.~Lake,
   {\it A New Approach to Generalised Uncertainty Relations},
   to appear in Touring the Planck Scale: Antonio Aurilia Memorial Volume, P. Nicolini ed., Springer, 
   [arXiv:2008.13183 [gr-qc]] (2020).
   
\bibitem{Lake-Frontiers-1}
   M.~J.~Lake, M.~Miller, R.~Ganardi and T.~Paterek,  
   {\it Generalised Uncertainty Relations from Finite-Accuracy Measurements},
   Front. Astron. Space Sci. \textbf{10}, 1087724 (2023).
   
   
\bibitem{Paunkovic:2022flx}
   N.~Paunkovic and M.~Vojinovic,
   {\it Equivalence Principle in Classical and Quantum Gravity},
   Universe \textbf{8}, 598 (2022).
   
\bibitem{Rae}
   A.~I.~M.~Rae, 
   {\it Quantum Mechanics}, 4th  ed.,
   Taylor \& Francis:  London, U.K. (2002). 
   
\bibitem{Gupta:2022qoq}
   R.~S.~Gupta, J.~Jaeckel and M.~Spannowsky,
   {\it Probing Poincar\'e Violation},
   [arXiv:2211.04490 [hep-ph]].
   
\bibitem{Amelino-Camelia:2014gga}
   G.~Amelino-Camelia,
   {\it Planck-scale soccer-ball problem: a case of mistaken identity},
   Entropy \textbf{19}, no.8, 400 (2017).
   
\bibitem{PerezdelosHeros:2022izj}
   C.~P\'erez de los Heros and T.~Terzi\'c,
   {\it Cosmic searches for Lorentz invariance violation},
   [arXiv:2209.06531 [astro-ph.HE]].

\bibitem{Kempf:1994su} 
   A.~Kempf, G.~Mangano and R.~B.~ Mann, 
   {\it Hilbert space representation of the minimal length uncertainty relation},
   Phys. Rev. D \textbf{52}, 1108-1118 (1995).

\bibitem{Kempf:1996ss} 
   A.~Kempf, 
   {\it On quantum field theory with nonzero minimal uncertainties in positions and momenta},
   J. Math. Phys. \textbf{38}, 1347-1372 (1997).
   
\bibitem{Ish95}
   C.~J.~Isham, 
   {\it Lectures on Quantum Theory: Mathematical and Structural Foundations},
   Imperial College Press, London, U.K. (1995). 

\bibitem{Nicolini:2010dj}
   P.~Nicolini and B.~Niedner,
   {\it Hausdorff dimension of a particle path in a quantum manifold},
   Phys. Rev. D \textbf{83}, 024017 (2011).
   
\bibitem{Lake:2022hzr}
   M.~J.~Lake,
   {\it Fractal properties of particle paths due to generalised uncertainty relations},
   Eur. Phys. J. C \textbf{82}, no.10, 928 (2022).

\bibitem{Moller:1959bhz}
   C.~M{\o}ller, 
   {\it The energy-momentum complex in general relativity and related problems},
   Colloq. Int. CNRS \textbf{91}, 15-29 (1962).
   
\bibitem{Rosenfeld:1963}
   L.~Rosenfeld, 
   {\it On quantization of fields}, 
   Nucl. Phys. 40, 353 (1963).
   
\bibitem{Kelvin:2019esx}
   Kelvin, K.~Onggadinata, M.~J.~Lake and T.~Paterek, 
   {\it Dark energy effects in the Schr{\" o}dinger-Newton approach},
   Phys. Rev. D \textbf{101}, no.6, 063028 (2020).

\bibitem{Nicolini:2012eu}
   P.~Nicolini, 
   {\it Nonlocal and generalized uncertainty principle black holes},
   [arXiv:1202.2102 [hep-th]] (2012).
 
\bibitem{Frankel:1997ec} 
   T.~Frankel, 
   {\it The Geometry of Physics: An Introduction},
   Cambridge University Press, Cambridge, U.K. (1997). 
     
\bibitem{Nakahara:2003nw} 
   M.~Nakahara, 
   {\it Geometry, Topology and Physics}, 2nd ed.,
   Taylor \& Francis, Boca Raton, FL, USA (2003).

\bibitem{Pikovski:2011zk}
   I.~Pikovski, M.~R.~Vanner, M.~Aspelmeyer, M.~S.~Kim and C.~Brukner,
   {\it Probing Planck-scale physics with quantum optics},
   Nature Phys. \textbf{8}, 393-397 (2012).

\bibitem{Bosso:2016ycv}
   P.~Bosso, S.~Das, I.~Pikovski and M.~R.~Vanner,
   {\it Amplified transduction of Planck-scale effects using quantum optics},
   Phys. Rev. A \textbf{96}, no.2, 023849 (2017).

\bibitem{Kumar:2017cka}
   S.~P.~Kumar and M.~B.~Plenio,
   {\it Quantum-optical tests of Planck-scale physics},
   Phys. Rev. A \textbf{97}, no.6, 063855 (2018).
   
\bibitem{Girdhar:2020kfl}
   P.~Girdhar and A.~C.~Doherty,
   {\it Testing generalised uncertainty principles through quantum noise},
   New J. Phys. \textbf{22}, no.9, 093073 (2020).

\bibitem{Cui:2021gsf}
   D.~Cui, T.~Li, J.~Li and X.~Yi,
   {\it Detecting deformed commutators with exceptional points in optomechanical sensors},
   New J. Phys. \textbf{23}, no.12, 123037 (2021).
      
\bibitem{Sen:2021oqj}
   S.~Sen, S.~Bhattacharyya and S.~Gangopadhyay,
   {\it Probing the generalized uncertainty principle through quantum noises in optomechanical systems},
   Class. Quant. Grav. \textbf{39}, no.7, 075020 (2022).
   

\bibitem{Marletto:2017pjr}
   C.~Marletto and V.~Vedral,
   {\it Why we need to quantise everything, including gravity},
   npj Quantum Inf 3, 29 (2017). 

\bibitem{Lake:2018zeg} 
   M.~J.~Lake, M.~Miller, R.~Ganardi, Z.~Liu, S.~D.~Liang and T.~Paterek,   
   {\it Generalised uncertainty relations from superpositions of geometries}.
   Class. Quant. Grav. \textbf{36}, no.15, 155012 (2019).

\bibitem{Lake:2019nmn}
   M.~J.~Lake, M.J, M.~Miller and S.~D.~Liang, 
   {\it Generalised uncertainty relations for angular momentum and spin in quantum geometry},
   Universe \textbf{6}, no.4, 56 (2020).
   
\bibitem{Perlmutter1999} 
   S.~Perlmutter {\it et al.}, 
   {\it Measurements of $\Omega$ and $\Lambda$ from 42 high-redshift supernovae}, 
   Astrophys. J. \textbf{517}, 565 (1999).
   
\bibitem{Reiss1998} 
   A.~G.~Riess {\it et al.}, 
   {\it Observational Evidence from Supernovae for an Accelerating Universe and a Cosmological Constant}, 
   Astron. J. \textbf{116}, 1009 (1998).
   
\bibitem{Giacomini:2017zju} 
   F.~Giacomini, E.~Castro-Ruiz and C.~Brukner, 
   {\it Quantum mechanics and the covariance of physical laws in quantum reference frames},
   Nature Commun. \textbf{10}, no.1, 494 (2019).
   
\bibitem{Lake:2020chb}
   M.~J.~Lake, 
   {\it Why space could be quantised on a different scale to matter},
   SciPost Phys. Proc. \textbf{4}, 014 (2021).
   
\bibitem{Lake:2021beh}
   M.~J.~Lake,
   {\it How Does the Planck Scale Affect Qubits?},
   Quantum Rep. \textbf{3}, no.1, 196-227 (2021).
   
\bibitem{Lake:2021gbu}
   M.~J.~Lake,
   {\it Generalised Uncertainty Relations and the Problem of Dark Energy},
   Romanian Astron. J. , Vol. 32, No. 1, p. 3–14, Bucharest (2022) 
   [arXiv:2112.13938 [gr-qc]].
 
\bibitem{Sahoo2004}
   D.~Sahoo, 
   {\it Mixing quantum and classical mechanics and uniqueness of Planck's constant}
   Journal of Physics A: Mathematical and General, Volume 37, Number 3 (2004).
   
\bibitem{Deser:2022lmi}
   S.~Deser,
   {\it Why even source-free gravity must be quantized},
   Eur. Phys. J. C \textbf{82}, no.5, 424 (2022).
 
\bibitem{LakeUkraine2019}
   M.~J.~Lake,
   {\it A Solution to the Soccer Ball Problem for Generalized Uncertainty Relations}.
   Ukrainian J.  Phys. \emph{64}, 1036 (2019).
                     
\end{thebibliography}
\end{document}